\documentclass[twocolumn,prl,showpacs,superscriptaddress,preprintnumbers,amsmath,amssymb]{revtex4}

\usepackage{graphicx}
\usepackage{dcolumn}
\usepackage{bm}
\usepackage{color}
\usepackage[latin1]{inputenc}
\usepackage[hang]{footmisc}

\begin{document}


\title{Resonant vibrations, peak broadening and noise in single molecule contacts: beyond the resonant tunnelling picture} 
\author{Daniel Secker}
\affiliation{Lehrstuhl für Angewandte Physik,
Friedrich-Alexander-Universität Erlangen-Nürnberg,\\Staudtstr. 7,
91058 Erlangen, Germany}
\author{Stefan Wagner}
\affiliation{Lehrstuhl für Angewandte Physik,
Friedrich-Alexander-Universität Erlangen-Nürnberg,\\Staudtstr. 7,
91058 Erlangen, Germany}
\author{Stefan Ballmann}
\affiliation{Lehrstuhl für Angewandte Physik,
Friedrich-Alexander-Universität Erlangen-Nürnberg,\\Staudtstr. 7,
91058 Erlangen, Germany}
\author{Rainer Härtle}
\affiliation{Institut für Theoretische Festkörperphysik,
Friedrich-Alexander-Universität Erlangen-Nürnberg,\\Staudtstr. 7,
91058 Erlangen, Germany}
\author{Michael Thoss}

\altaffiliation{Interdisziplinäres Zentrum für Molekulare
Materialien, Friedrich-Alexander-Universität Erlangen-Nürnberg}

\affiliation{Institut für Theoretische Festkörperphysik,
Friedrich-Alexander-Universität Erlangen-Nürnberg,\\Staudtstr. 7,
91058 Erlangen, Germany}
\author{Heiko B. Weber}

\email[]{heiko.weber@physik.uni-erlangen.de}

\altaffiliation{Interdisziplinäres Zentrum für Molekulare
Materialien, Friedrich-Alexander-Universität Erlangen-Nürnberg}

\affiliation{Lehrstuhl für Angewandte Physik,
Friedrich-Alexander-Universität Erlangen-Nürnberg,\\Staudtstr. 7,
91058 Erlangen, Germany}





\date{\today}

\begin{abstract}

We carry out experiments on single-molecule junctions at low
temperatures, using the mechanically controlled break junction
technique. Analyzing the results received with more than ten different molecules
the nature of the first peak in the differential
conductance spectra is elucidated. We observe an electronic
transition with a vibronic fine structure, which is most
frequently smeared out and forms a broad peak. In the usual
parameter range we find strong indications that additionally
fluctuations become active even at low temperatures. We conclude
that the electrical field feeds instabilities, which are triggered
by the onset of current. This is underscored by noise measurements
that show strong anomalies at the onset of charge transport.

\end{abstract}


\maketitle

\newpage

We consider charge transport across single conjugated molecules.
The fundamental picture to understand the current voltage
characteristics is the resonant tunnelling model
\cite{Datta,LindsayRatner}. At low bias, no charge transport
occurs due to the absence of a resonant level and Coulomb
blockade. In lowest order, current starts to flow at a certain
threshold, at which the electrons are energetically allowed to
occupy an orbital in the molecule. The consequence is a step-like
increase of the current $I$($V$), or correspondingly a peak in the
differential conductance d$I$/d$V$($V$). In the framework of a
pure electronic transition, it is well known that the peak width
of the differential conductance is determined by the temperature
$T$ and the coupling constant $\Gamma$. We have measured more than
ten molecular species and none of them showed correspondingly
narrow peaks at the onset of conductance, even at low temperatures
(significantly smaller than 10\,mV, in contrast to the observed
100-300\,mV). Hence there is a qualitative mismatch between the
purely electronic resonant tunnelling model and experimental data:
the peaks appear by far too broad.

This is partially resolved when vibrations are considered
\cite{Galperin07}. Their impact on $IV$ characteristics is
expected to be inelastic sidepeaks to the electronic peaks. Such
signatures have been resolved in rather few experiments, where the
first peak is explicitly tuned towards low bias voltages ($<
100$\,meV) by an external gate
\cite{ParkPark,QiuNazinHo,OsorioRubenZant}, only in rare cases
without an external gate \cite{SBallmann}. However, it is not
obvious how many and which vibrational modes are excited and show
up as additional transport channels in the differential
conductance. Equidistant peaks have been observed \cite{SBallmann}
which indicate a fundamental vibrational excitation and its higher
harmonics. Theory shows that a single vibronic mode can trigger a
bunch of excitations \cite{HaertleThossPRL}. Additionally, dynamic
behavior of electron transport is not only driven by vibronic
excitations, but also by fluctuation processes due to
instabilities in the contact configuration \cite{SeckerWeber}.  In
the following we show that the resonant tunnelling model
(including vibrations) is suitable for the description of the
first peak only at small voltages. We show further deviations from
this model, which are associated to bias-induced fluctuations.

\begin{figure}
    \centering
        \includegraphics[width=\columnwidth]{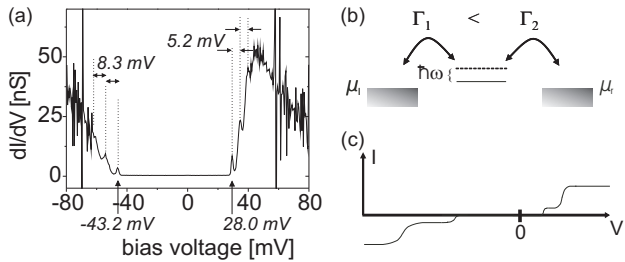}
                        \caption{(a) d$I$/d$V$ measurement at the first electronic peak for a small blockade region. Several single excitations can be recognized, for positive voltages the distance is $\approx$\,5.2\,mV, for negative $\approx$\,8.3\,mV. The onset of current shows the same asymmetry, taking place at $-43.2$\,mV and at $+28.0$\,mV. (b) Schematic representation of two leads with chemical potentials $\mu_{\rm{l}}$ and $\mu_{\rm{r}}$ coupled asymmetrically ($\Gamma_{\rm{1}}$\,$>$\,$\Gamma_{\rm{2}}$) to a molecular level, which is accompanied by a vibrational state $\hbar \omega$. (c) Resulting current-voltage characteristic. The onset of current and peak distance is lower for positive voltages at the weaker coupled left electrode (in case of LUMO-transport).}
    \label{fig:vibEnergie}
\end{figure}

\begin{figure}[h]
    \centering
        \includegraphics[width=\columnwidth]{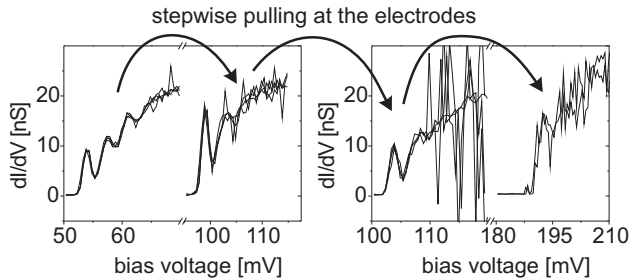}
                \caption{d$I$/d$V$ measurements at the onset of current for a small blockade region (left curve) and after pulling at the electrodes in three subsequent steps. With increasing onset voltage the vibrational features smear out step by step.}
    \label{fig:Ziehen}
\end{figure}

\begin{figure}[h]
    \centering
        \includegraphics[width=0.75\columnwidth]{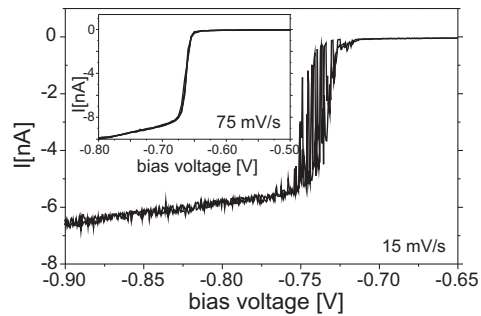}
        \caption{$I$-$V$ measurements that show high fluctuations at the onset of current when lowering the scanning speed from 75\,mV/s (inset) by a factor of five.}
    \label{fig:Current_Onset}
\end{figure}

\begin{figure}
    \centering
        \includegraphics[width=0.8\columnwidth]{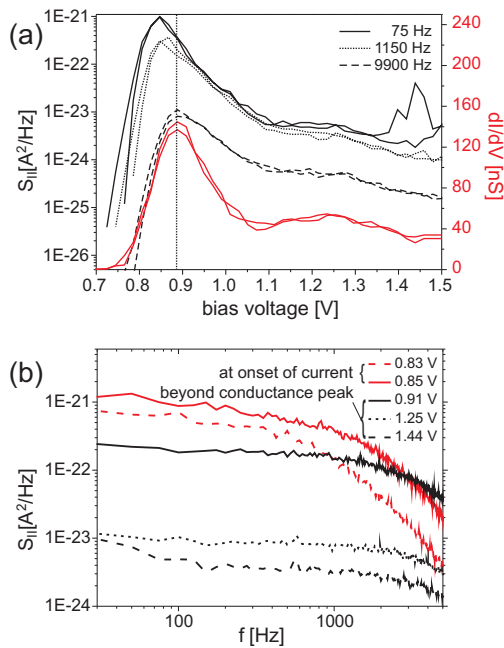}
        \caption{(a)Spectral noise density at three different frequencies and d$I$/d$V$ curve (red) at the first conductance peak. At the onset of current the noise is several orders of magnitude enhanced. (b) Noise spectra at the onset of current (red) and beyond the conductance peak (black) show different frequency dependence ($f^{-2}$ vs. $f^{-1}$) above 1\,kHz.}
    \label{fig:Noise}
\end{figure}

We use the mechanically controlled break junction (MCBJ)
technique. The method is desribed elsewhere \cite{Reichert}. The
features discussed here have been found to be general for several ($>10$)
organic compounds. They all have a delocalized $\pi$-electron
system and a length of about 2\,nm, some of them containing at
least one metal ion center. As an example, the molecule
Fe\textsuperscript{2+}-bis(pyrazolyl)pyridine \cite{Ruben} is used for the
measurements presented in the figures \ref{fig:vibEnergie},
\ref{fig:Ziehen} and \ref{fig:Current_Onset} and an
oligo(phenylene vinylene) (OPV-3) molecule in the figure
\ref{fig:Noise}.

We first turn our attention to the low-voltage regime. Typically
the first conductance peak occurs at some hundred millivolts.
However, this value varies substantially from junction to junction
as a consequence of the uncontrolled local environment
\cite{Reichert}. In few junctions, the onset of current occurs at
few tens of millivolts. Here, the expectations from the resonant
tunnelling picture are well fulfilled and a closer look elucidates
the physics of the first peak.

A measurement of current and voltage in figure
\ref{fig:vibEnergie}, taken with a resolution of 0.5\,mV per step
reveals a substructure in the general peak form.
Several peaks appear on a background of a rather large peak
structure.

We analyze the equidistant fine structure and find that the peak
separation at negative and positive bias voltages has a ratio of
$8.3/5.2$\,$=$\,1.6. Correspondingly, the onset of current has the
same ratio of $\left|-43.2/28.0\right|$\,$=$\,1.6. This can
readily be explained by a molecular electronic level that is
accompanied by vibronic levels at intervals of $\hbar \omega$ and
coupled asymmetrically to the electrodes, as illustrated in Fig.
\ref{fig:vibEnergie}b. The asymmetry creates asymmetric voltage
drops at the two molecule-metal interfaces. So in one voltage
direction the molecular levels will be in resonance at smaller
absolute voltages, and also the peak separation will be larger if
the stronger coupled electrode moves along the levels. This leads
to a current-voltage characteristic as plotted in Fig.
\ref{fig:vibEnergie}c. The onset of current and the peak
separation at both polarities fulfill the same relation as the
coupling to the electrodes, if the voltage drop is assumed to
satisfy the same ratio.

A vibrational excitation energy of $E_{\text{vib}}\approx3.2$\,meV
can be extracted.
This qualitatively fits to the spectrum of vibrational modes that
are obtained from DFT-calculations (BP86/def-TVZP) for the free molecule.
For example, there are thirteen modes with frequencies
in the range of $0.8$\,meV to $10$\,meV.

\begin{figure}
    \centering
        \includegraphics[width=0.75\columnwidth]{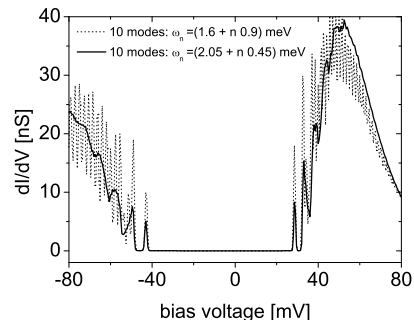}
                        \caption{$\text{d}I/\text{d}V$ characteristics obtained
from a nonequilibrium Green's function
calculation for a resonant level model \cite{HaertleThossPRL,Hartle,Hartle10}.
The model comprises an electronic
state $39$\,meV above the system's Fermi-energy and ten
vibrational modes with frequencies $\omega_{n}$ that are coupled to this
state with a uniform coupling strength such that the reorganization energy amounts
$22$\,meV for both plots.
In the calculation,  the vibrational modes are assumed to be in their thermal
equilibrium state determined by the junction's temperature $T=1$\,K.
The level-width functions $\Gamma_{1}=13$\,$\mu$eV and
$\Gamma_{2}=9$\,$\mu$eV have been chosen to model the current and voltage
drop retrieved from figure \ref{fig:vibEnergie}a.}
    \label{fig:Theorie}
\end{figure}

In principle, the observed peak shape can be modelled within the
resonant tunnelling model, assuming electronic-vibrational
coupling to a multitude of vibrational modes. As an example,
figure \ref{fig:Theorie} shows results of nonequilibrium Green's
function model calculations \cite{HaertleThossPRL,Hartle,Hartle10}, which
reproduce the experimental data in figure \ref{fig:vibEnergie}a
rather well. The underlying model includes a single electronic
state on the molecular bridge, located $39$\,meV above the
Fermi-energy, which is coupled uniformly to ten harmonic
vibrational modes (with a corresponding reorganisation energy of
$22$\,meV for both plots). The frequencies of the vibrational
modes are chosen as $\omega_{n}=2.05+n\,0.45$\,meV and
$\omega_{n}=1.6+n\,0.9$\,meV for the black and the red line,
respectively ($n\in\{1..10\}$).

In agreement with the experimental data, the model calculations
predict a broad peak in the conductance. It is emphasized that the width of this peak
is not related to molecule-lead coupling ($\Gamma_{1/2}\approx0.01$\,meV) or
thermal broadening ($k_{\text{B}}T\approx0.1$\,meV) but is a result of the
coupling to the distribution of vibrational modes
and is approximately determined by the overall reorganization energy of $22$\,meV.
A more detailed analysis of the conductance reveals a complex substructure,
the details of which depend on the specifics of the model \cite{Hartle}. The
first narrow peak appears once the chemical potential
in one of the leads allows the population of the resonant level.
At larger bias voltages the single conductance-peaks
overlap with each other and form a broad peak. The small gap between the
first narrow peak and the following broadened structures is determined
by the frequency $\omega_{1}$.
It is noted that vibrational relaxation processes, e.g. due to coupling
to phonons in the gold contacts, as well as anharmonic effects,
which are neglected in the present calculation, will
result in a further broadening of the peak structures.

Although the calculations within the resonant tunnelling model
show that the observed peak structure can be explained in
principle, our experiments give also information that supports a
more complex physics. With the MCBJ technique, we are able to
stretch stable junctions. Microscopically, we expect that the gold
tip is further pulled out, which results either in a deformation
of the contact bonds and/or in a reduction of the self capacitance
of the molecular junction. Experimentally, this goes along with an
enlargement of the blockade region in d$I$/d$V$($V$). Hence, the
first peak occurs at higher voltages. When the junction is further
stretched, a sudden reorganisation of the contact or a disruption
may occur. We recorded current-voltage characteristics at four
successive positions (see Fig. \ref{fig:Ziehen}). Considering the
resolution of vibronic excitations, in the first measurement three
distinct excitations can be resolved, the second shows only two of
them, in the third only the first appears clearly while the fourth
does not reveal a clear substructure anymore.  Since only the
blockade region is enlarged and the junction is essentially
unaltered in the process, the reason for the smearing of the
vibronic excitations should be the larger electric field. The
change in electrode spacing is estimated to about 13\,pm, and can
be neglected compared to the length of the molecule of 2\,nm. Only
a slight modification is seen in the first vibronic peak that
changes the amplitude relative to the others, but the increasing
smearing towards higher bias is the most obvious effect. Within
the model, this can be assigned to a change in the
electron-vibration coupling parameters. However, the nearly
ubiquitous observation of smeared-out peaks at higher bias is
rather likely to be explained by an additional smearing mechanism.

At this point, a detailed analysis of the first peak with respect
to fluctuations is desirable. We show in Fig.
\ref{fig:Current_Onset} a relatively sharp onset of current in the
high bias regime ($\approx -700$\,mV). An apparently stable
current-voltage characteristic is observed at a measurement speed
of the voltage sweep of 75\,mV/s (inset). Repeating the
measurement at a slower speed of 15\,mV/s in this case (not
always) leads to high current fluctuations. We conclude that there
are instabilities of the IV characteristics, which are associated
to structural reconfigurations, given the slow timescales.

In order to get a deeper understanding of the timescales of these
fluctuations noise measurements are carried out parallel to the
$I$-$V$-characteristics at a similar molecular contact. Fig.
\ref{fig:Noise}a shows the spectral noise density as a function of
voltage for three chosen frequencies as well as the differential
conductance (red curve) at the first peak with an FWHM of
$\approx$\,150\,mV. The current noise is low in the blockade
region, and increases by four orders of magnitude close to the
maximum of the conductance peak. It then decreases again by two
orders of magnitude after the first peak. This is a finding of
high importance for the understanding of the first peak, and is
not anticipated by the resonant tunnelling models. A closer look
reveals that the maximum of the noise is actually at lower bias
than the maximum of the conductance peak, in particular at lower
frequencies. In other words, the fluctuations are maximum when the
current starts to flow. This reminds a model proposed by Koch et
al. \cite{Koch}. The authors considered an increased Fano factor
(i.e. shot noise level) in the weak coupling regime, with a
Franck-Condon weighted electron-vibration coupling. The basic idea
is that in the transport model tunnelling is energetically allowed
when the first electronic level is reached, but suppressed due to
a weak wave function overlap (Franck-Condon blockade). When
reaching a bias which allows also to excite a vibration, the
molecule starts to vibrate as a consequence of a first tunnelling
event, and under some realistic assumptions a stronger overlap
will allow more electrons to tunnel through the junction. In their
model, avalanche-like amplification of the current occurs, and
with a certain probability the current breaks down to zero after a
while. These statistics leads to giant shot noise signals at the
onset of current flow. Although the experimental situation is more
complex than this theoretical and, in addition, is limited to shot
noise rather than configurational noise, we think that some
essential ideas are also valid for single-molecule junctions in
the high-bias regime. We consider the case that the onset of
current triggers configurational fluctuations. One may suspect
that there exist field-sensitive bistabilities, which need a
certain field strength to be active. The onset of current and
eventually of first vibrational excitations may be the trigger of
further field-induced structural reconfigurations, which appear as
low-frequency noise or instabilities. Alternatively, a deformation
of the molecule might act on the threshold value itself, providing
a situation similar to a polaron instability
\cite{Bratkovsky,Ryndyk,Galperin07}. As we know, the average
charge on the molecule changes when the current sets in. This may
induce a structural change, or may create structural
bistabilities, also on low timescales. Common to these
illustrative considerations is the observed phenomenon that the
onset of current triggers reconfigurations, accompanied by large
fluctuations. When further increasing the bias the system
undergoes a strongly fluctuating regime, with first a maximum in
the noise signal and subsequently a maximum in d$I$/d$V$. Beyond
the two maxima, the current reaches a higher level (similar to the
resonant tunnelling model), whereas the noise is reduced. This
indicates an approximately steady state. The noise measurements
indicates that the physics of the first peak is much more
complicated than its similarity to the resonant tunnelling model
suggests.

Some further information can be extracted from the frequency
dependence of the current noise density (Fig. \ref{fig:Noise}b).
The presented curves at five different voltages split into two
pairs, the red curves at the onset of current, and the
black ones for bias voltages beyond the maximum of the
differential conductance. The first pair approximately follows a
decay $\propto$\,$f^{-2}$, whereas the rest resembles more to a
$f^{-1}$-dependence. While the higher exponent is ascribed by a
simple Lorentz oscillator model to a single fluctuator
\cite{Imry}, the latter can be explained by the superposition of
several oscillators with uniformly distributed time constants
\cite{OchsSecker, Mueller}.

Hence, there are two different origins of noise identified. On the
one hand the first electronic transition is accompanied by one
dominating bistability of the molecular contact configuration.
This was surprising to us and may point to an intrinsic effect,
for example the polaron model, rather than to fluctuations in the
gold leads. Beyond the peak, many fluctuators are active. Hence,
the first current plateau should be understood as a dynamic
interplay of charge flow and dynamical reconfigurations of the
molecule. This picture goes well beyond the resonant tunnelling
model.

Altogether, from many experiments with more than ten conjugated
molecules and from detailed calculations, we receive the following
picture of charge transport through single molecules: The purely
electronic transmission is accompanied by distinct vibrational
resonant conduction channels. At low voltage, these smear out due
to finite temperature and for a broad distribution of vibronic
modes leading to a rather broad peak. More complicated physics
happens at even higher electric field. The transition from the
blockade regime to the conducting regime shows strong
configurational fluctuations, with a noise maximum before the
maximum of the differential conductance. At these higher bias
thresholds, as soon as current flows in a molecule, strong
structural fluctuations are associated to it. Hence the similarity
of the $IV$ characteristics to the resonant tunnelling model may
be rather fortuitous at higher bias.

\begin{acknowledgments}
We acknowledge funding of the DFG in the framework of the SPP 1243
\textquotedblleft Quantum transport on the molecular
scale\textquotedblright. The work was carried out in the
 Cluster of Excellence \textquotedblleft Engineering of Advanced
Materials\textquotedblright at the Friedrich-Alexander-Universität
Erlangen-Nürnberg.

\end{acknowledgments}


\begin{thebibliography}{99}



\bibitem{Datta} S. Datta, \textit{Electronic transport in mesoscopic
systems}, Cambridge 1995.

\bibitem{LindsayRatner}
S. M. Lindsay and M. A. Ratner, \textit{Adv. Mater.} ,
\textbf{19}, 23 (2007).

\bibitem{Galperin07}
M. Galperin, M. A. Ratner and A. Nitzan, \textit{J. Phys.:
Condens. Matter} \textbf{19}, 103201 (2007).

\bibitem{ParkPark} H. Park, J. Park, A. K. L. Lim, E. H. Anderson, A. P. Alivisatos and P. L. McEuen,
\textit{Nature} \textbf{407}, 57 (2000).

\bibitem{QiuNazinHo} X. H. Qiu, G. V. Nazin and W. Ho,
\textit{Phys. Rev. Lett.} \textbf{92}, 206102 (2004).

\bibitem{OsorioRubenZant} E. A. Osorio, M. Ruben, J. S. Seldenthuis, J.-M. Lehn and H. S. J. van der Zant,
\textit{small} \textbf{6}, 174 (2010).

\bibitem{SBallmann} S. Ballmann, W. Hieringer, D. Secker, Q. Zheng, J. A. Gladysz, A. G\"orling and H. B. Weber,
\textit{ChemPhysChem} \textbf{11}, 2256 (2010).

\bibitem{HaertleThossPRL} R. H\"artle, C. Benesch and M. Thoss,
\textit{Phys. Rev. Lett.} \textbf{102}, 146801 (2009).

\bibitem{SeckerWeber} D. Secker and H. B. Weber,
\textit{Phys. Stat. Sol.} \textbf{11}, 4176 (2007).




\bibitem{Reichert} J. Reichert, R. Ochs, D. Beckmann, H. B. Weber, M. Mayor and H. v. L\"ohneysen,
\textit{Phys. Rev. Lett.} \textbf{88}, 176804 (2002).
\bibitem{Ruben}  C. Rajadurai, F. Schramm, O. Fuhr, M. Ruben, \emph{Eur. J. Inorg. Chem}., \textbf{17} 2649 (2008); and I. Salitros, N.T. Madhu, R. Boca, J. Pawlik, M. Ruben, \emph{Monatsh.
Chem.} \textbf{140}, 695 (2009).

\bibitem{Hartle} R. H\"artle, C. Benesch and M. Thoss, \textit{Phys. Rev. B} \textbf{77}, 205314 (2008).
\bibitem{Hartle10} R. H\"artle, R. Volkovich, M. Thoss and U. Peskin, \textit{J. Chem. Phys.} \textbf{133}, 081102 (2010).
\bibitem{Koch} J. Koch and F. von Oppen, \emph{Phys. Rev. Lett.} \textbf{94},
206804 (2005).
\bibitem{Bratkovsky} A. S. Alexandrov, and A. M. Bratkovsky,
\textit{Phys. Rev. B} \textbf{67}, 235312 (2003).

\bibitem{Ryndyk} D. A. Ryndyk, P. DAmico, G. Cuniberti, and K. Richter \textit{Phys. Rev. B} \textbf{78}, 085409 (2008).


\bibitem{Imry} Y. Imry,
\textit{Introduction to mesoscopic physics} \textbf{1997}, \textit{Oxford University Press}, New York.

\bibitem{OchsSecker} R. Ochs,  D. Secker, M. Elbing, M. Mayor, and H.B. Weber, \textit{Faraday Discussions} \textbf{131}, 281 (2006).

\bibitem{Mueller} R. M\"uller,
\textit{Rauschen} \textbf{1990}, \textit{Axel Springer Verlag}, Berlin.




\end{thebibliography}
\end{document}